\documentclass{article}

\usepackage[main,nonatbib,final]{neurips_2026}

\usepackage[utf8]{inputenc} 
\usepackage[T1]{fontenc}    
\usepackage{hyperref}       
\usepackage{url}            
\usepackage{booktabs}       
\usepackage{multirow}       
\usepackage{rotating}       
\usepackage{amsfonts}       
\usepackage{nicefrac}       
\usepackage{microtype}      
\usepackage[table,dvipsnames]{xcolor} 
\usepackage{algorithm}
\usepackage{algpseudocode}
\usepackage{amsmath}
\usepackage{siunitx}
\usepackage{float}
\usepackage{marvosym}
\usepackage{wrapfig}
\usepackage{pifont}
\usepackage{environ}  
\usepackage{graphicx}
\usepackage{verbatim}

\newif\ifshowdraft
\showdraftfalse  

\usepackage[dvipsnames]{xcolor}
\NewEnviron{draft}{%
  \ifshowdraft
    {\color{red} \BODY}
  \fi
}

\newcommand{\norm}[1]{\left\lVert#1\right\rVert}
\def\X{{\mathbf X}}
\def\d{{\mathrm d}}
\def\W{{\mathbf W}}

\newcommand{\Xt}[1]{\X_{t_{#1}}}
\newcommand{\muforward}[1]{\Tilde{\mu}_{k}}
\newcommand{\mubackward}[1]{\mu_{k+1}}
\newcommand{\sigmaforward}[1]{\Tilde{\sigma}^2_{k}}
\newcommand{\sigmabackward}[1]{\sigma^2_{k+1}}

\title{SE-MSB: End-to-End Unpaired Speech Enhancement using Mamba Schrödinger Bridges}

\author{
  Andreas Bagge$^{*\text{{\Cross}}}$, Andreas Nymand$^{* \text{{\Cross}}}$, Michael Riis Andersen$^*$ and Bjørn Sand Jensen$^*$  \\  
   $^{*}$DTU Compute, Kgs. Lyngby, Denmark \\
  $^\text{{\Cross}}$ WSA, Lynge, Denmark \\
  \{andreas.bagge,andreas.nymand\}@wsa.com , \{miri, bjje\}@dtu.dk   
}

\begin{document}

\begin{draft}
    Draft notes:
    \begin{itemize}
        \item Why do we use a Mamba model architecture? 
        Why Mamba2 when there is a Mamba3?
        \item Explain the metrics. Why those metrics?
        Are there any up/down sides to those metrics?
        Are there any "missing" metrics?
        
    \end{itemize}
\end{draft}

\maketitle

\begin{abstract}
%
%
%
%

Speech enhancement (SE) models typically rely on supervised learning with paired data examples where clean speech is synthetically degraded. 
This paradigm limits performance in real-world scenarios where the target environment's specific acoustic characteristics are unknown. 
We propose a fully unpaired SE framework that uses principled Diffusion Schrödinger Bridges (DSB) to learn a stochastic transport process between a clean and a degraded speech distribution.
Algorithms for learning transport maps are computationally heavy since they require simulating differential equations during training, usually at each training step.
Therefore, we propose using a high-efficiency Mamba Diffusion Model designed for end-to-end waveform processing. 
We compare against state-of-the-art methods for speech enhancement, both paired and unpaired, as well as a classical signal processing algorithm.
Experimental results show that we are on par or better than the baselines while being orders of magnitude faster during inference. 
Furthermore, we show that the flexibility of the DSB formulation allows our model to generalize across SE tasks, offering a robust and efficient solution for real-world speech restoration.


\end{abstract}

\section{Introduction}
Speech enhancement (SE) problems, such as denoising, declipping, or dereverberation, are classic problems in audio processing. Traditionally, these are solved using supervised methods trained on collections of paired instances of clean speech $x_0$ and degraded speech  $x_1$, where the degraded speech is typically simulated by modifying the clean speech, e.g., by adding noise or clipping the amplitude such that $x_1 \sim p_\text{noisy}(\cdot|x_0)$.
Such approaches are limited to training on simulated distortions and cannot be applied to actual, in-the-wild degraded audio because large-scale paired data collection is infeasible or, in some instances, even impossible. 
Enabling model training on unpaired in-the-wild audio, i.e., where clean and degraded speech are drawn from separate distributions $x_0 \sim p_\text{clean}, x_1 \sim p_\text{noisy}$ independently rather than paired samples $(x_0, x_1) \sim p_{\text{joint}}$, has the 
potential to improve SE models in real-world conditions by adapting the model to out-of-distribution (OOD) data, for example, by personalizing the model to the user's specific acoustic environment.
With the recent rise in personalized AI, especially through federated learning~\cite{kulkarni2020surveypersonalizationtechniquesfederated}, the ability to train models directly on observed, but unpaired or unlabeled, data is becoming increasingly important for adapting models to the specific needs of end-users.
This is especially true for SE models, where the end-user's acoustic environment is often unknown and can be highly variable and unpredictable. 
In some instances, the degradation is not only unknown but also difficult to simulate to be used in a paired setup, e.g., Lombard effect~\cite{lombard} or reconstructing historical records~\cite{lemercier2024diffusionmodelsaudiorestoration}.
Diffusion models constitute the state-of-the-art for image and audio synthesis~\cite{dhariwal2021diffusionmodelsbeatgans, kong2021diffwaveversatilediffusionmodel}, 
and they have also shown great results for paired SE~\cite{sgmse}. Diffusion models generally consist of two processes, a forward and a backward process. 
In traditional diffusion models, the forward process gradually turns data samples into Gaussian noise, and the objective of the model is to learn the backward process by effectively removing noise from the input. 
After learning the reverse process, the models can then synthesize high-quality samples from Gaussian noise. 
Though traditional diffusion methods yield high-quality samples, they do not allow for mapping between two arbitrary data distributions.
In contrast, Schrödinger Bridges (SB)~\cite{Schrödinger1932} allow for high-quality sample generation and direct transfer between two arbitrary data distributions through the entropy-regularized optimal transport problem, also called the SB problem, of finding the most likely random evolution between two continuous probability distributions. 
That is, SBs allow learning stochastic maps between clean and degraded audio from unpaired data only, enabling training on in-the-wild data. SBs are formulated in terms of stochastic differential equations (SDEs), which can be simulated with SDE solvers, and solving the SB problem therefore allows for a diffusion-based generative approach.
Mamba is a selective state-space model that serves as an efficient alternative to the
popular transformer architecture for long sequence modelling~\cite{mamba2}. Mamba offers linear scaling in sequence length
and contains a constant state size during inference. 
We propose the SE-MSB (Speech Enhancement Mamba Schrödinger Bridges) model, which is a novel combination of Diffusion Schrödinger Bridges (DSB) and modern state-space models for end-to-end SE using completely unpaired data.
We provide a fast and efficient implementation of the DSB method for speech enhancement on raw audio waveforms using a bespoke Mamba Diffusion model. Raw waveforms suffer from a higher dimensionality, which is a problem solved by Mambas linear sequence length scaling. 
We show that the combination of DSB and Mamba outperforms or is on par with other state-of-the-art methods for unpaired SE, while being orders of magnitude faster during inference, both due to the efficiency of the network architecture and due to the ability of the model to do few-step diffusion without sacrificing performance.
Audio examples and code are available at \href{a}{anonymous.4open.science/r/Latent-DSB}. Our main contributions are: 

\begin{figure}[t]
    \centering
    \includegraphics[width=0.7\linewidth]{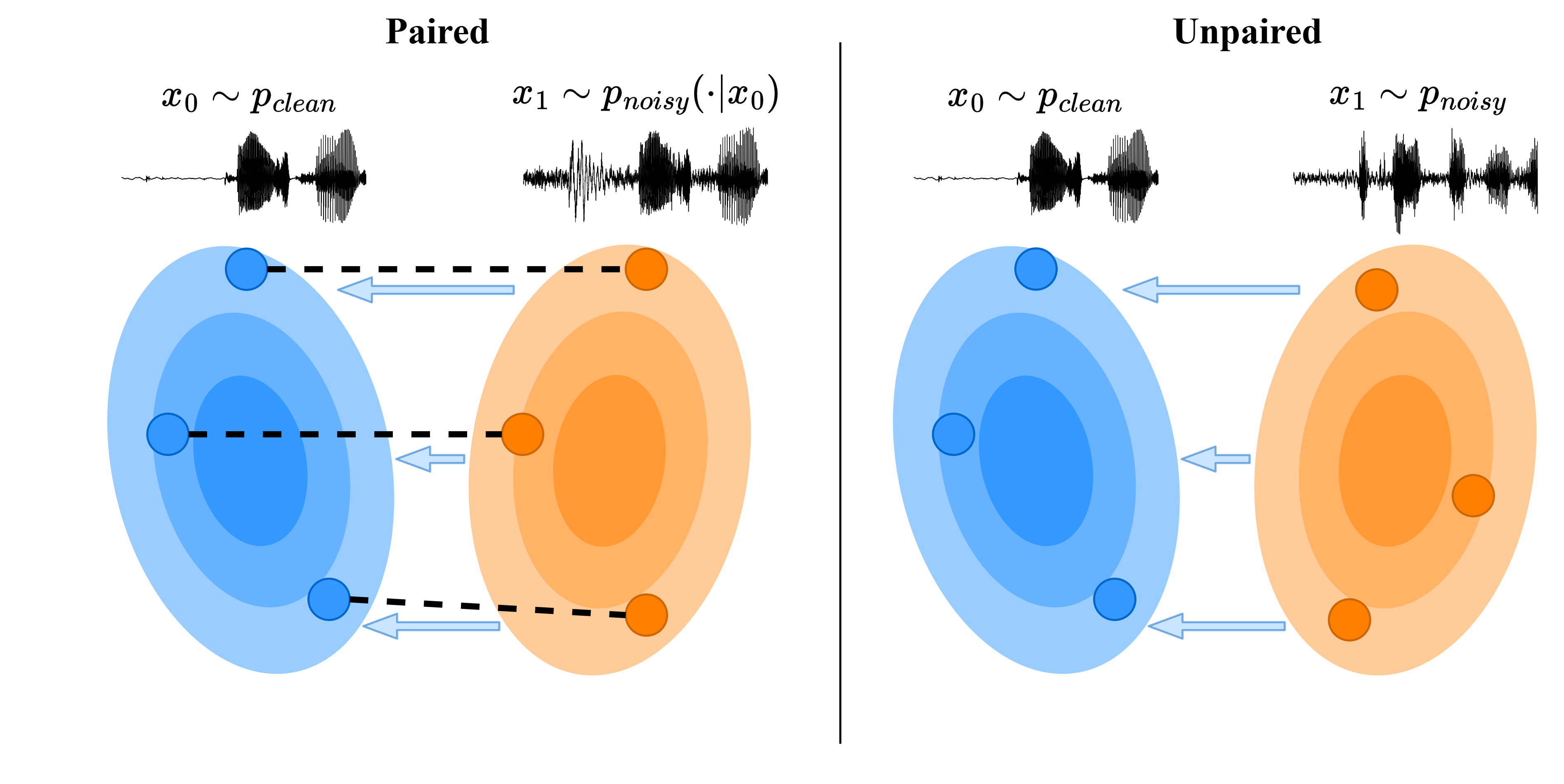}
    \vspace{-0.5cm}
    \caption{In traditional supervised speech enhancement (\textbf{left}), 
    we assume access to a clean sample $x_0$ from the \textcolor{blue}{clean target distribution} and a corresponding degraded sample from the \textcolor{orange}{degraded distribution} to learn the optimal pairing (blue arrows). 
    In contrast, our method (\textbf{right}) requires only independent samples from both distributions to learn the optimal pairing.}
    \label{fig:placeholder}
\end{figure}

{
\setlength{\leftmargini}{1.5em}
\setlength{\parskip}{0pt}
\setlength{\parsep}{0pt}
\begin{itemize}     
  \setlength{\itemsep}{0pt}
  \setlength{\parskip}{0pt}
  \setlength{\parsep}{0pt}
    \item We propose the SE-MSB model, which is the \textbf{first fully unpaired and end-to-end model} for speech enhancement on raw waveforms.
    \item We evaluate SE-MSB using a rigorous experimental protocol and compare it against several state-of-the-art baselines on \textbf{multiple tasks and compounded conditions}.  
    \item We demonstrate that SE-MSB achieves \textbf{strong speech enhancement performance}, outperforming paired methods in some cases while exhibiting \textbf{highly competitive efficiency}. 
\end{itemize}
}

\section{Related Work}
A$^2$ASB (Audio-to-Audio Schrödinger Bridge) \cite{kong2025a2sbaudiotoaudioschrodingerbridges} is a tractable, 
simulation-free method for learning maps between the distributions $p_{data}(\X_0)$ and $p_{prior}(\X_1|\X_0)$, 
i.e., a conditional probability distribution that requires paired samples. 
A$^2$ASB exhibits state-of-the-art results on paired audio restoration tasks such as bandwidth extension and inpainting. 
CycleGAN is a discriminative generative model for learning a deterministic mapping between distributions~\cite{zhu2020unpairedimagetoimagetranslationusing}. CycleGANs have also been used for style transfer in the context of audio, for example, for whisper-to-normal speech conversion~\cite{patel2020cincganeffectivef0prediction}. However, CycleGANs are highly unstable during training and require jointly optimizing 4 distinct neural networks. 

\newcommand{\yes}{{\color{green}\ding{52}}}
\newcommand{\no}{{\color{red}\ding{54}}}
\begin{wraptable}{r}{0.6\textwidth}
  \centering
  \begin{tabular}{l c c c}
    \toprule
     & Unpaired & End-to-End & Task-flexible \\
    \midrule
    SE-MSB      & \yes & \yes & \yes \\
    GFB     & \yes & \no $^*$ & \yes \\
    BUDDy    & \yes   & \no & \no \\
    A$^2$ASB & \no  & \no & \yes \\
    \bottomrule
  \end{tabular}
  \caption{Comparison of current paired/unpaired diffusion models for speech enhancement.
  $^*$GFB processes raw audio but employs an internal STFT encoder and ISTFT decoder.}
\end{wraptable}

\cite{moliner2024gaussianflowbridgesaudio} proposes using Gaussian Flow Bridges for unpaired speech enhancement, 
a diffusion-based approach where samples are first mapped to an intermediate Gaussian distribution.
The approach suffers from the fact that the simulated trajectories first have to map to a third distribution, a Gaussian, therefore yielding non-optimal transport. 
A similar class of methods for unpaired speech enhancement uses diffusion posterior sampling to 
solve the audio inverse problem in an unsupervised manner \cite{hernandezolivan2023vrdmgvocalrestorationdiffusion,buddy,moliner2023solvingaudioinverseproblems}. 
Instead of training a task-specific model, these methods utilize a pretrained diffusion model as a prior over the clean data.
During inference, the models then modify the reverse diffusion process to guide the sampling trajectory.
However, these methods also require access to the mathematical structure of the degradation process,
which, in most cases, is not available for in-the-wild data. 
Diffusion Schrödinger Bridge Matching \cite{sb_flow, shi2023diffusionschrodingerbridgematching} is an algorithm for computing the Schrödinger Bridge, 
a dynamic entropy-regularised version of optimal transport. 
The algorithm can be used for learning a diffusion model for mapping between arbitrary distributions using unpaired data samples.
Mamba, a selective state-space model architecture, has shown great results in the field of paired speech enhancement~\cite{zhang2025mambaspeechalternativeselfattention} \cite{yang2026schrodingerbridgemambaonestep}. In combination with diffusion Schrödinger bridges, Mamba has shown to outperform Long Short-Term Memory and Multi-Head Self-Attention backbones for Short Time Fourier Transform (STFT) based audio representations for paired speech enhancement. However, STFT-based methods are limited by fixed transformation hyperparameters, reducing the model's ability to capture the expressivity of raw audio.


\section{Method}
Our work builds upon the Diffusion Schrödinger Bridge Matching algorithm \cite{sb_flow, shi2023diffusionschrodingerbridgematching}, DSB for short.
Consider two distinct continuous probability distributions on $\mathbb{R}^n$. 
In the literature, these distributions are usually denoted as $p_{data}$ and $p_{prior}$.
However, for our application of speech enhancement, we will denote these distributions as $p_{clean}$ and $p_{noisy}$, respectively, to emphasize their role as the target and source distributions in the speech enhancement problem.
Therefore, $p_{clean}$ refers to the distribution of clean speech, while $p_{noisy}$ refers to the distribution of degraded speech, 
for example, noisy, reverberant, or clipped speech.
We will denote samples from $p_{clean}$ and $p_{noisy}$ as $\X_0$ and $\X_1$, respectively, i.e. $\X_0\sim p_{clean}$ and $\X_1\sim p_{noisy}$, 
where $\X_0,\X_1\in\mathbb{R}^n$.
We seek a stochastic process, or a 'bridge', that transports samples from $p_{clean}$ to $p_{noisy}$ and vice versa. 
We will parameterize the processes as two distinct SDEs, one for each direction, such that simulating the SDEs with a
sample as the initial value will transport said sample from one domain to the other. We will denote the forward and backward SDEs as:
\begin{align*}
    \d\X_t&=f(\X_t, t)\d t+g(t)\d\W_t, \hspace{5mm} \X_0\sim p_{clean}, \\
    \d\X_t&=b(\X_t, t)\d t+g(t)\d\bar{\W}_t, \hspace{5mm} \X_1\sim p_{noisy},
\end{align*}
where $f$ and $b$ are the drift terms, $g(t)$ is the diffusion term, and $\W$ and $\bar{\W}$ are standard Wiener processes. 
For simplicity, we will assume $g(t)=\beta$ is constant. 
Therefore, the problem reduces to learning the drift terms for each process, i.e., $f$ and $b$, such that simulating the backward process from $t=1$ to $t=0$ starting from $\X_1 \sim p_{noisy}$ yields $\X_0 \sim p_{clean}$ at time $t=0$ (and vice versa for the forward process).
We will parameterize the forward and backward drift with a single neural network $v(\X_t, t, s)$ with parameters $\theta$ such that:
\begin{equation*}
    f(\X_t,t)\approx v_{\theta}(\X_t,t,1),\hspace{5mm}b(\X_t,t)\approx v_{\theta}(\X_t,t,0),
\end{equation*}
where $s \in\{0,1\}$ is a binary indicator variable, indicating whether the forward or backward drift is being approximated. Alternatively, 
one could have parameterized each drift with a separate neural network.
Having obtained the optimal parameters $\theta^*$, we can approximately simulate the forward and backward SDEs on the discrete time interval 
$0=t_0<t_1\dots < t_N=1$, where $\Delta t_{k+1}=t_{k+1}-t_k$, using the Euler-Maruyama method:
\begin{equation}
    \begin{aligned}
        p(\Xt{k+1}|\Xt{k}) &\approx p(\Xt{k+1}|\Xt{k},\X_1) =\mathcal{N}(\Xt{k+1};\muforward{},\sigmaforward{}\mathbf{I}), \hspace{5mm} \text{(forward process)}
    \end{aligned}
    \label{eq:p_forward}
\end{equation}
\begin{equation}
    \begin{aligned}
        p(\Xt{k}|\Xt{k+1}) &\approx p(\Xt{k}|\Xt{k+1},\X_{0})=\mathcal{N}(\Xt{k};\mubackward{},\sigmabackward{}\mathbf{I}). \hspace{5mm} \text{(backward process)}
    \end{aligned}
    \label{eq:p_backward}
\end{equation}
That is, the forward and backward processes can be simulated by sampling from the respective Gaussian distributions, 
where the means and variances are given by:
\begin{equation*}
    \begin{aligned}
        \muforward{} &\approx \Xt{k}+\Delta t_{k+1}\underbrace{(\X_1-\Xt{k})/(1-t_k)}_{=f(\Xt{k}, t_k)} \approx \Xt{k}+\Delta t_{k+1} v_{\theta}(\Xt{k},t_k,1), \\
        \mubackward{} &\approx \Xt{k+1}+\Delta t_{k+1}\underbrace{(\X_0-\Xt{k+1})/t_{k+1}}_{=b(\Xt{k+1}, t_{k+1})} \approx \Xt{k+1}+\Delta t_{k+1}v_{\theta}(\Xt{k+1},t_{k+1},0),
    \end{aligned}
\end{equation*}
\begin{gather*}
    \sigmaforward{} = \frac{\beta\Delta t_{k+1}(1-t_{k+1})}{1-t_k}, \hspace{5mm} \sigmabackward{} = \frac{\beta\Delta t_{k+1} t_k}{t_{k+1}}.
\end{gather*}
Additionally, it can be shown~\cite{sb_flow} that given $\X_0$ and $\X_1$, 
then intermediate points in the bridge can be sampled from the conditional distribution:
\begin{equation}
    p(\Xt{k}|\X_0,\X_1)=\mathcal{N}((1-t_k)\X_0+t_k\X_1,\beta t_k(1-t_k)\mathbf{I}).
    \label{eq:intermediate}
\end{equation}
To train the neural network $v_{\theta}$, we will use the DSB Matching algorithm \cite{sb_flow, shi2023diffusionschrodingerbridgematching}.  
The algorithm contains two phases: pre-training and fine-tuning. 
During the pre-training phase, i.i.d. samples from $p_{clean}$ and $p_{noisy}$ are sampled independently. We will call this pair a "random coupling" (i.e., unpaired). An intermediate point, $\Xt{k}$, is sampled from the conditional 
distribution \eqref{eq:intermediate}, and the objective of the network is to predict the forward and backward
drift at time $t_k$ and $t_{k+1}$ given by $(\X_1 - \Xt{k})/(1-t_k)$ 
and $(\X_0 - \Xt{k+1})/t_{k+1}$, respectively. After a number of pre-training steps, the fine-tuning phase begins. 
Here, the initial samples $\X_0$ and $\X_1$ are once again obtained by sampling from $p_{clean}$ and $p_{noisy}$, but they are no longer 
randomly coupled. Instead, the couplings are obtained by simulating the processes on the interval from $t=0$ to $t=1$ and from $t=1$ to $t=0$, 
respectively, using the current parameters $\theta$ of the model. The objective of the model is the same as in the pre-training phase.
\cite{sb_flow} shows that applying the pre-training phase followed by the fine-tuning phase will eventually
yield the SB, i.e., the optimal stochastic process between $p_{clean}$ and $p_{noisy}$.
The DSB Matching algorithm can be seen as a stochastic version of the Reflow algorithm \cite{liu2022flowstraightfastlearning}, 
which can be deduced from DSB Matching by setting $\beta=0$, i.e., by removing the noise from the SDEs. 
For the Reflow algorithm, the authors show that the fine-tuning phase yields "straight flows", 
i.e. flows that transport samples from $p_{clean}$ to $p_{noisy}$ along straight lines.
This is a desirable property for diffusion models, as it allows for few-step diffusion sampling without compromising sample quality, therefore requiring less compute during inference.
Additionally, it can also be shown that A$^2$ASB is simply a paired version of the DSB Matching algorithm that uses only the pre-training phase, and that only trains the backward process, but instead of drawing random couplings, the couplings are drawn according to the joint distribution and are therefore paired, see Appendix \ref{appendix:baselines}. 

\newcommand{\iid}[1]{\overset{\mathrm{iid}}{\sim}}
\definecolor{darkgreen}{RGB}{0,100,0} 

\begin{algorithm}[t]
    \caption{Training algorithm}
    \label{box:training}
    \begin{algorithmic}[1]
        \State Initialize neural network $v_{\theta}$ and batch size $2B$ 
        \State Let $i \in \{ 1,2\dots B \}$
        \While{not converged}
            \If{Pre-training}
                \State $\X_0^{1:B}\iid{} p_{clean}$, $\Tilde{\X}_1^{1:B}\iid{} p_{noisy}$
                \State $\X_1^{1:B}\iid{} p_{noisy}$, $\Tilde{\X}_0^{1:B}\iid{} p_{clean}$
            \Else
                \State $\X_0^{1:B}\iid{} p_{clean}$, $\Tilde{\X}_1^{1:B}\iid{} p_{noisy}$
                \State Sample $\X_1^i$ using \eqref{eq:p_forward} and $v_{\theta}(\cdot, \cdot, 1)$ starting from $\X_0^i$
                \State Sample $\tilde{\X}_0^i$ using \eqref{eq:p_backward} and $v_{\theta}(\cdot, \cdot, 0)$ starting from $\tilde{\X}_1^i$
            \EndIf
            \State $t^{1:B}, \Tilde{t}^{1:B} \iid{} \mathcal{U}(0,1)$
            \State $\X_t^i \sim \mathcal{N}((1-t^i)\X_0^i + t^i\X_1^i, \beta t^i(1-t^i)\mathbf{I})$
            \State $\Tilde{\X}_t^i \sim \mathcal{N}((1-\Tilde{t}^i)\tilde{\X}_0^i + \Tilde{t}^i\tilde{\X}_1^i, \beta \Tilde{t}^i(1-\Tilde{t}^i)\mathbf{I})$
            \State $\mathcal{L}_b = \frac{1}{B}\sum_{i=1}^B \norm{v_{\theta}(\X_t^i,t^i,0) - \frac{\X_0^i-\X_t^i}{t^i}}$
            \State $\mathcal{L}_f = \frac{1}{B}\sum_{i=1}^B \norm{v_{\theta}(\tilde{\X}_t^i,\Tilde{t}^i,1) - \frac{\tilde{\X}_1^i-\tilde{\X}_t^i}{1-\Tilde{t}^i}}$
            \State Take gradient step $\nabla_{\theta}\frac{1}{2}(\mathcal{L}_b+\mathcal{L}_f)$
        \EndWhile
    \end{algorithmic}
\end{algorithm}

\subsection*{Mamba Diffusion Model architecture}
\begin{figure}[ht]
    \centering
    \includegraphics[width=1.0\linewidth]{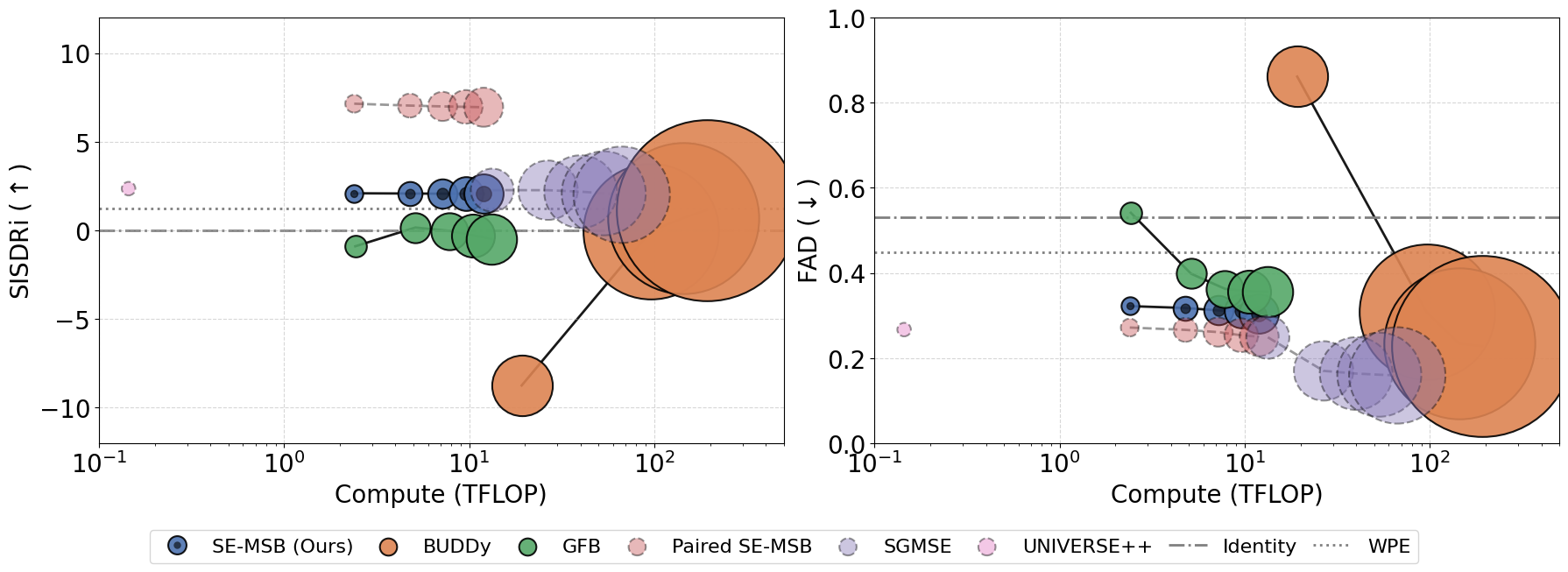}
    \label{fig:compute_vs_metrics}
    \caption{Total compute vs SISDRi \textbf{(left)} and FAD \textbf{(right)}. The area of the points scales linearly with mean inference time, see table \ref{tab:compute} and appendix \ref{appendix:efficiency}. Our model outperforms unpaired state-of-the-art models with the same compute budget. It is also seen that the model is robust to changes in the number of diffusion steps, performing almost the same no matter how many diffusion steps are taken. This is also shown in figure \ref{fig:diffusion_steps}. The models with a dashed outline are trained on paired data. An Equivalent plot for the CHiME-6 dataset can be seen in appendix \ref{appendix:efficiency}.}
\end{figure}
To the best of our knowledge, all existing unpaired speech enhancement models operate on spectrograms
\cite{hernandezolivan2023vrdmgvocalrestorationdiffusion,buddy,moliner2023solvingaudioinverseproblems}. 
This is done mostly to enable the use of established convolutional neural networks, such as U-Nets~\cite{ronneberger2015unetconvolutionalnetworksbiomedical}, on the resulting time-frequency representations. 
However, such time-frequency representations impose a rigid, non-learnable trade-off between temporal and frequency resolution, and, in the case of magnitude representations, completely disregard the phase, therefore requiring heuristic phase reconstruction algorithms or an additional vocoder for waveform synthesis. 
Therefore, we propose combining DSB with modern state-space architectures to operate in an end-to-end fashion on raw audio waveforms, enabling the model to learn an internal representation of the raw data, without relying on specific choices of time-frequency representations.
To this end, we use a bespoke, efficient Mamba Diffusion model architecture based on Mamba blocks \cite{mamba2}.
This architecture uses an initial 1D convolutional layer to learn an internal latent representation of the raw audio waveform, similar to a learnable time-frequency representation. 
This latent is then processed by a number of Mamba Diffusion blocks. 
Each Mamba Diffusion block uses Adaptive LayerNorm \cite{xu2019understandingimprovinglayernormalization} 
to inject the diffusion timestep $t$ and the process-direction conditioning $c$. 
The model architecture is visualized in Figure~\ref{fig:model_architecture}, 
and a more detailed description of the architecture can be found in Appendix~\ref{appendix:model_description}.
To compare the algorithmic efficiency of the proposed SE-MSB model to the state-of-the-art baselines, we have opted to count tera ($10^{12}$) floating-point operations (TFLOPS) required to evaluate a sample output. Each of the models processed a dummy audio input of $2^{15} = 32768$ samples (2.048 seconds of audio at 16kHz). For counting the TFLOPs of the models, PyTorch's "FlopCounterMode" was used. As our implementation relied on the \texttt{mamba-ssm} library, which uses custom Triton kernels for an efficient implementation, the TFLOP count has been manually registered and is counted as explained in chapter 6 of the Mamba2 paper~\cite{mamba2}. The majority of the Mamba2 TFLOP count comes from the 1D convolution and the selective scan algorithms, which, according to the paper, consist of an order $O(TN^2)$ FLOPs, where $T$ is the sequence length and $N$ is the state dimension. This is under the assumption that $N = P = Q$ where $P$ is the head dimension and $Q$ is the chunk size. If we assume a more general form, it can be shown that the FLOP count from the selective scan equals $FLOP_{scan} = 2 B T  N(H\ Q + 3 D)$, where $B$ is the batch size, $H$ is the number of heads, and $D$ is total head dimension, $D = H  P$, see appendix \ref{appendix:efficiency}. Before the block decomposition, a 1D causal convolution across the inner dimension is performed with a total cost of $FLOP_{conv} = 2 B  T D  K$, where $K$ is the convolution kernel size. In addition to this, some elementwise operations are used in the Triton kernels, and are, as such, invisible to Torch's "FlopCounterMode". The FLOP count is negligible compared to the selective state scan, but for completeness, we include them in the count. First, the SiLU activation layers take approximately $5 B T  D$ FLOPs. Secondly, the gated RMS-Norm takes $11 B T  D$ FLOPs, 5 from the RMS norm, 5 for an attached SiLU, and 1 from an elementwise multiplication. Lastly, a final elementwise multiplication takes $BT H  Q$ FLOPs.
Further information on the FLOP count on non-torch baselines can be seen in Appendix \ref{appendix:efficiency}.
\begin{figure}
    \centering
    \includegraphics[width=0.7\linewidth]{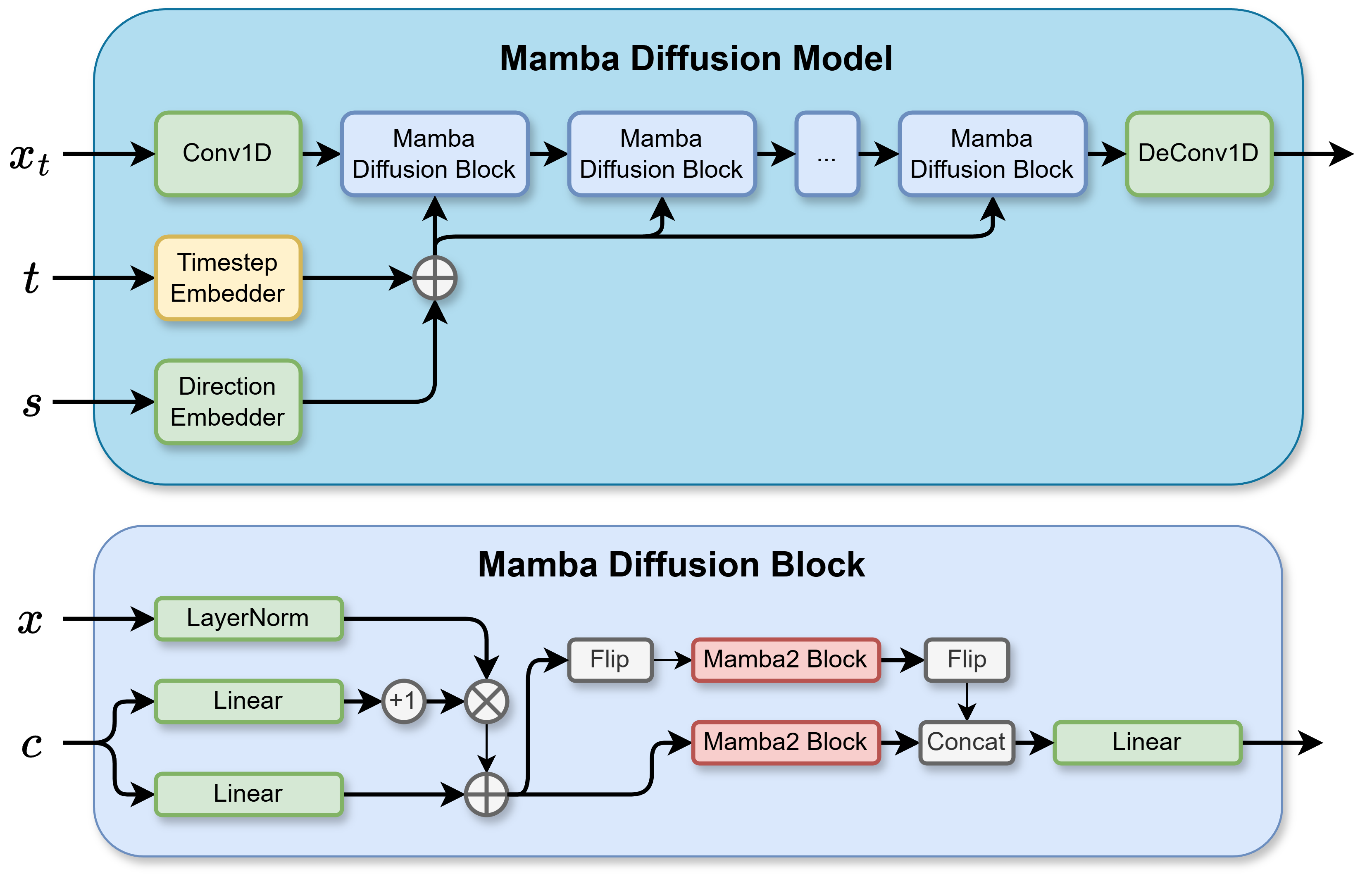}
    \label{fig:model_architecture}
    \caption{The Mamba Diffusion Model architecture proposed for SE-MSB. Each waveform is encoded using a 1D convolutional layer. The resulting representation is mixed with the conditional signal and processed by a stack of Mamba Diffusion Blocks.}
\end{figure}
\vspace{-3mm}

\section{Experiments}
The proposed SE-MSB model is trained and evaluated on a dereverberation, denoising, and declipping task.
These tasks are traditionally solved with supervised methods, but by artificially degrading the clean speech data, we can evaluate distortion metrics that require paired data.
The clean speech data used throughout our experiments comes from VCTK~\cite{vctk}, see Appendix~\ref{appendix:datasets}. 
The VCTK dataset is both used as $p_{clean}$ and to simulate $p_{noisy}$, but, importantly, the model never sees pairs of clean and degraded audio during training.

\begin{figure}[t]
\centering
\includegraphics[width=0.9\textwidth]{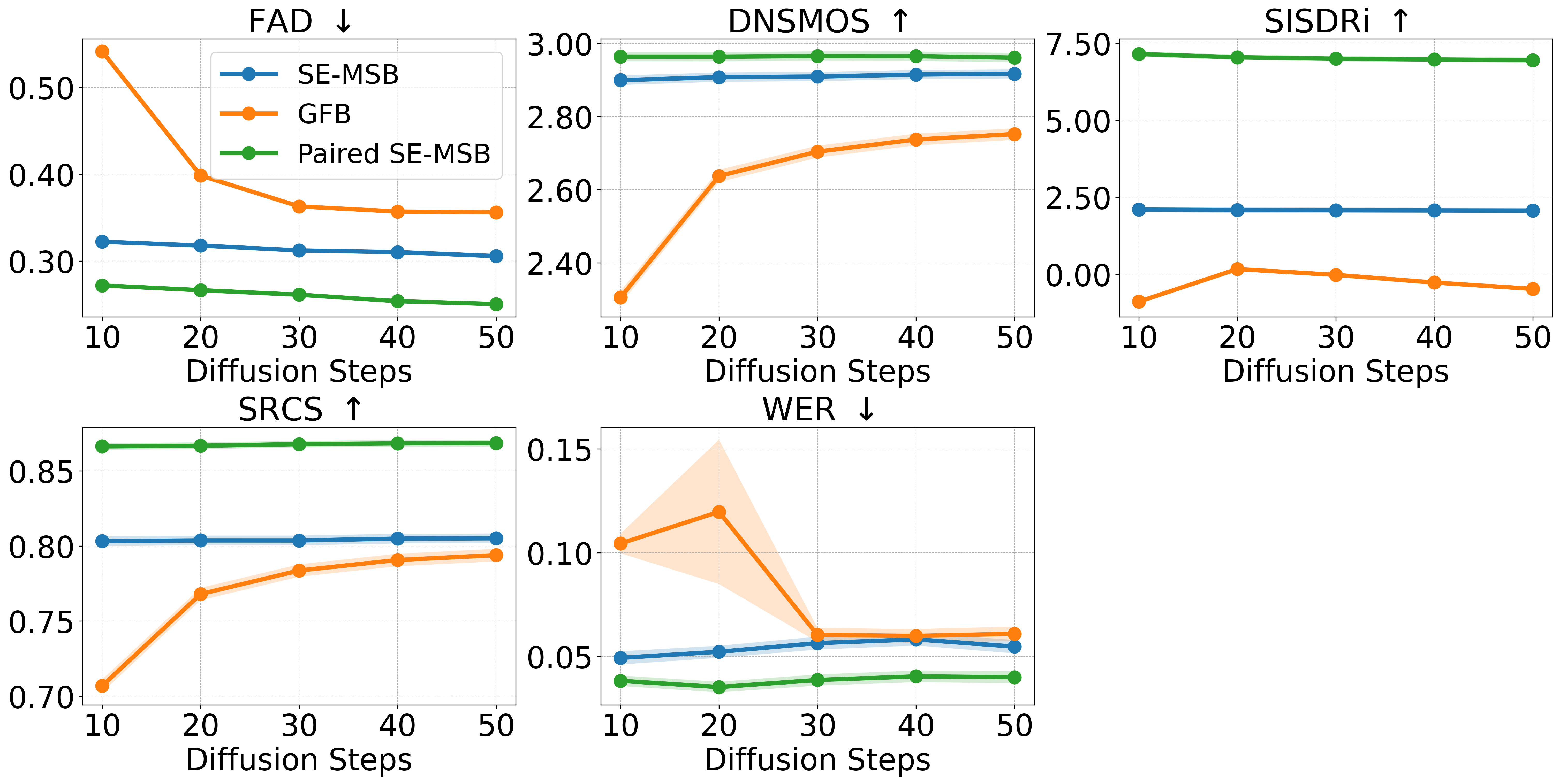}
\caption{
    Metrics as a function of the number of diffusion steps. The SE-MSB model is almost unaffected by the number of steps, 
    showing that SE-MSB can do few-step generation. 
    }
\label{fig:diffusion_steps}
\end{figure}

The models are tested on the test split of VCTK using an unseen test set of room impulse responses. Standard supervised speech enhancement models are conventionally evaluated using isolated test sets to assess generalizability to OOD shifts. In contrast, the primary advantage of unpaired speech enhancement is the ability for in-domain training, allowing the model to train directly on the observed target distribution. Therefore, to verify that our model retains generalizability, we also evaluate all models on a separate, independent clean speech dataset, namely the Libri ASR Corpus~\cite{libri}.
{
\setlength{\leftmargini}{1.5em}
\setlength{\parskip}{0pt}
\setlength{\parsep}{0pt}
\begin{itemize}     
  \setlength{\itemsep}{0pt}
  \setlength{\parskip}{0pt}
  \setlength{\parsep}{0pt}
    \item \textbf{Dereverberation:} The degraded speech is generated by applying room impulse responses (RIRs) 
    from various public RIR datasets~\cite{mitreverb, butreverb, openairreverb, rwcpreverb, c4dmreverb} to the clean speech.
    \item \textbf{Denoising and declipping:} The degraded speech is generated via additive noise
    and, in a separate experiment, clipping the resulting mixture. 
    The noise dataset is WHAM! noise~\cite{wichern2019whamextendingspeechseparation}
    mixed at 5 dB SNR. Clipping is randomly applied to remove between 0 and 1 dB of the signal power.
\end{itemize}
}
\begin{table}[t]
\centering
\renewcommand{\arraystretch}{1.2} 
\resizebox{\textwidth}{!}{%
\begin{tabular}{@{} c l l S[table-format=1.2] 
  S[table-format=2.2 \pm 2.2, separate-uncertainty=true] 
  S[table-format=-2.2 \pm 2.2, separate-uncertainty=true] 
  *{2}{S[table-format=2.2 \pm 2.2, separate-uncertainty=true]} 
  S[table-format=3.4] @{}}
\toprule
  & & \textbf{Model} & {\textbf{FAD ($\downarrow$)}} & {\textbf{DNSMOS ($\uparrow$)}} & {\textbf{SISDRi ($\uparrow$)}} & {\textbf{WER ($\downarrow$)}} & {\textbf{SR-CS ($\uparrow$)}} & {\textbf{TFLOPS ($\downarrow$)}} \\
\midrule
\rowcolor{gray!20} \cellcolor{white} & U & SE-MSB (10)     & 0.32      & 2.90 +- 0.01        & {\textbf{2.09 $\pm$ 0.09}}        & 0.05 +- 0.00          & 0.80 +- 0.00 & 2.3937 \\
\rowcolor{gray!20} \cellcolor{white} & U & SE-MSB (50)     & 0.31      & 2.92 +- 0.01        & 2.06 +- 0.09        & 0.05 +- 0.00          & 0.81 +- 0.00 & 11.9687 \\
\rowcolor{gray!20} \cellcolor{white} & U & BUDDy (10 $\dagger$)    & 0.86    & 2.30 +- 0.01    & -8.75 +- 0.11    & 0.54 +- 0.01      & 0.47 +- 0.01 & 19.2127 \\
\rowcolor{gray!20} \cellcolor{white} & U & BUDDy (100 $\dagger$)   & {\textbf{0.23}}  & {\underline{\textbf{3.08 $\pm$ 0.01}}}  & 1.15 +- 0.12  & {\underline{\textbf{0.02 $\pm$ 0.00}}}    & {\underline{\textbf{0.91 $\pm$ 0.00}}} & 192.1273 \\
\rowcolor{gray!20} \cellcolor{white} & U & GFB (10)     & 0.54      & 2.30 +- 0.02        & -0.90 +- 0.08        & 0.10 +- 0.00          & 0.71 +- 0.00 & 2.4265 \\
\rowcolor{gray!20} \cellcolor{white} & U & GFB (50)     & 0.36      & 2.75 +- 0.02        & -0.48 +- 0.04        & 0.06 +- 0.00          & 0.79 +- 0.00 & 13.2112 \\
\rowcolor{gray!20} \cellcolor{white} & U & WPE          & 0.45        & 2.71 +- 0.02            & 1.24 +- 0.07            & {\underline{\textbf{0.02 $\pm$ 0.00}}}              & 0.84 +- 0.00 & {\underline{\textbf{0.0012}}} \\
\cmidrule{2-9}
 & P & SGMSE (10)   & 0.25    & 2.89 +- 0.02    & 2.27 +- 0.09    & {\underline{0.02 $\pm$ 0.00}}      & 0.90 +- 0.00 & 13.3338 \\
 & P & SGMSE (50)   & {\underline{0.16}}    & 2.92 +- 0.02    & 2.01 +- 0.11    & {\underline{0.02 $\pm$ 0.01}}      & {\underline{0.91 $\pm$ 0.00}} & 66.6690 \\
 & P & Paired SE-MSB (10) $^{\star}$   & 0.27    & 2.96 +- 0.01    & {\underline{7.15 $\pm$ 0.08}}    & 0.04 +- 0.00      & 0.87 +- 0.00 & 2.3937 \\
 & P & Paired SE-MSB (50) $^{\star}$   & 0.25    & 2.96 +- 0.01    & 6.95 +- 0.08    & 0.04 +- 0.00      & 0.87 +- 0.00 & 11.9687 \\
 & P & UNIVERSE++   & 0.27    & 3.05 +- 0.01    & 2.35 +- 0.09    & 0.03 +- 0.00      & 0.90 +- 0.00 & 11.9687 \\
 \cmidrule{2-9}
\multirow{-12}{*}{\rotatebox[origin=c]{90}{\textbf{VCTK}}} & & Identity     & 0.53   & 2.64 +- 0.02  & 0.00 +- 0.00  & {\underline{0.02 $\pm$ 0.00}}    & 0.79 +- 0.00 & 0.0000 \\
\midrule
\rowcolor{gray!20} \cellcolor{white} & U & SE-MSB (10)     & {\underline{\textbf{0.15}}}      & 2.90 +- 0.02        & 0.60 +- 0.09        & 0.11 +- 0.02          & 0.67 +- 0.00 & 2.3937 \\
\rowcolor{gray!20} \cellcolor{white} & U & SE-MSB (50)     & 0.16      & 2.92 +- 0.02        & 0.52 +- 0.09        & 0.13 +- 0.01          & 0.66 +- 0.00 & 11.9687 \\
\rowcolor{gray!20} \cellcolor{white} & U & BUDDy (10 $\dagger$)    & 0.69    & 2.21 +- 0.02    & -9.95 +- 0.15    & 0.76 +- 0.02      & 0.38 +- 0.01 & 19.2127 \\
\rowcolor{gray!20} \cellcolor{white} & U & BUDDy (100 $\dagger$)   & 0.17   & {\underline{\textbf{3.18 $\pm$ 0.01}}}  & -1.76 +- 0.18  & 0.04 +- 0.00    & 0.80 +- 0.00 & 192.1273 \\
\rowcolor{gray!20} \cellcolor{white} & U & GFB (10)     & 0.31      & 2.59 +- 0.02        & -1.14 +- 0.07        & 0.16 +- 0.01          & 0.59 +- 0.01 & 2.4265 \\
\rowcolor{gray!20} \cellcolor{white} & U & GFB (50)     & 0.29      & 2.90 +- 0.02        & -0.33 +- 0.04        & 0.08 +- 0.00          & 0.71 +- 0.01 & 13.2112 \\
\rowcolor{gray!20} \cellcolor{white} & U & WPE          & 0.22        & 2.71 +- 0.03            & {\textbf{1.40 $\pm$ 0.07}}            & {\underline{\textbf{0.02 $\pm$ 0.00}}}              & {\textbf{0.89 $\pm$ 0.00}} & {\underline{\textbf{0.0012}}} \\
\cmidrule{2-9}
 & P & SGMSE (10)   & 0.25    & 2.89 +- 0.02    & 2.27 +- 0.09    & {\underline{0.02 $\pm$ 0.00}}      & 0.90 +- 0.00 & 13.3338 \\
 & P & SGMSE (50)   & 0.16    & 2.92 +- 0.02    & 2.01 +- 0.11    & {\underline{0.02 $\pm$ 0.01}}      & {\underline{0.91 $\pm$ 0.00}} & 66.6690 \\
 & P & Paired SE-MSB (10) $^{\star}$   & 0.20    & 3.01 +- 0.02    & {\underline{2.75 $\pm$ 0.12}}    & 0.08 +- 0.00      & 0.67 +- 0.01 & 2.3937 \\
 & P & Paired SE-MSB (50) $^{\star}$   & 0.20    & 3.03 +- 0.02    & 2.72 +- 0.12    & 0.08 +- 0.00      & 0.67 +- 0.01 & 11.9687 \\
 & P & UNIVERSE++  & 0.22    & 2.96 +- 0.02    & -1.08 +- 0.17    & 0.18 +- 0.02      & 0.66 +- 0.01 & 11.9687 \\
 \cmidrule{2-9}
\multirow{-12}{*}{\rotatebox[origin=c]{90}{\textbf{Libri}}} & & Identity     & 0.28   & 2.57 +- 0.03  & 0.00 +- 0.00  & 0.03 +- 0.00    & 0.82 +- 0.00 & 0.0000 \\
\bottomrule
\end{tabular}%
}
\caption{
    Evaluation metrics for the dereverberation task with 95\% confidence intervals. 
    The Identity baseline corresponds to unprocessed audio.
    The number of diffusion steps is denoted in parentheses when relevant, except for $\dagger$, which denotes optimization steps.
    U = unpaired, P = paired. Best unpaired model in each metric is \textbf{bold}, and best model across paired and unpaired models is \underline{underlined}.
}
\label{tab:reverberation}
\end{table}
\begin{table}[t]
\centering
\renewcommand{\arraystretch}{1.2}
\resizebox{0.8\textwidth}{!}{%
\begin{tabular}{c l 
    S[table-format=3.4, round-mode=places, round-precision=4]
    S[table-format=5.2 \pm 2.2, separate-uncertainty=true] 
    S[table-format=2.2]}
\toprule
& \textbf{Model} & {\textbf{TFLOPS}} & {\textbf{Inference Time (ms)}} & {\textbf{Trainable Params ($\times 10^6$)}} \\
\midrule
\rowcolor{gray!20} U & SE-MSB (10)             & 2.3937399999999998  & 214.19 +- 0.53 & 45.56 \\
\rowcolor{gray!20} U & SE-MSB (50)             & 11.9687  & 1042.29 +- 6.59 & 45.56 \\
\rowcolor{gray!20} U & BUDDy (10 $\dagger$)    & 19.21273 & 2481.45 +- 8.02 & 27.74 \\
\rowcolor{gray!20} U & BUDDy (100 $\dagger$)   & 192.12731 & 22135.4 +- 17.1 & 27.74 \\
\rowcolor{gray!20} U & GFB (10)                & 2.42654 & 314.67 +- 0.44 & 43.85 \\
\rowcolor{gray!20} U & GFB (50)                & 13.21116 & 1713.99 +- 3.99 & 43.85 \\
\rowcolor{gray!20} U & WPE                     & 0.0012 & 80.72 +- 0.25 & 0.0 \\
\midrule
P & Sepformer                                 & 0.47215 & 55.21 +- 0.18 & 25.61 \\
P & SGMSE (10)                                 & 13.3338 & 1253.28 +- 1.85 & 65.59 \\
P & SGMSE (50)                                 & 66.66901 & 6276.25 +- 5.7 & 65.59 \\
P & Paired SE-MSB (10)                              & 2.3937399999999998 & 215.0 +- 0.39 & 45.56 \\
P & Paired SE-MSB (50)                              & 11.9687 & 1031.56 +- 0.73 & 45.56 \\
P & UNIVERSE++                                      & 0.144606 & 124.95 +- 2.36 & 84.24 \\
\bottomrule
\end{tabular}%
}
\caption{
    Computational efficiency: FLOPS, inference time, and number of trainable parameters.
}
\label{tab:compute}
\end{table}
For each task, we train our SE-MSB model using the Mamba Diffusion Model architecture 
on raw audio waveforms using $\beta=0.1$ following a parameter sweep. 
The model is trained for 250k steps during pre-training and 100k steps during fine-tuning for a total of 350k steps.
We use a batch size of 18 on audio samples of 1.00 seconds at a sample rate of 16 kHz. During the fine-tuning phase, each SDE is simulated using 20 steps of the Euler-Maruyama method.
For dereverberation, we compare the performance of the SE-MSB model against both unpaired and paired baselines.
The unpaired baselines include a Gaussian Flow Bridge (GFB) method~\cite{moliner2024gaussianflowbridgesaudio},
a diffusion posterior sampling method called BUDDy~\cite{buddy},
and the Weighted Prediction Error (WPE) method\cite{wpe}. 
For the paired baselines, we evaluate against a Score-based Generative Model for Speech Enhancement (SGMSE)\cite{sgmse},
and the supervised version of the SE-MSB method called A$^2$SB\cite{kong2025a2sbaudiotoaudioschrodingerbridges}.
We also include an Identity baseline corresponding to unprocessed reverberant audio. Baselines are described in detail in Appendix~\ref{appendix:baselines}.

To showcase the flexibility of our method, we train an SE-MSB model on a mixture of multiple degradations: reverberation, additive noise, and clipping. 
We compare the performance of the SE-MSB model against BUDDy, and the unprocessed baseline.
BUDDy uses a diffusion posterior sampling method,
but BUDDy also assumes a specific degradation, namely a convolution with some unknown RIR, 
and is therefore sensitive to these assumptions being violated.
For the tasks involving additive noise, we also compare against a paired baseline, 
namely the Sepformer~\cite{sepformer}, a state-of-the-art supervised speech enhancement model.

We test the few-step sampling capabilities of both the paired and unpaired SE-MSB model compared to the unpaired diffusion baseline GFB
on the dereverberation task by evaluating metrics for different numbers of sampling steps.
The number of sampling steps can be changed at inference time, and therefore, only a single trained model is required for each task.
Few-step sampling is a desirable property as it can reduce the inference time of the model,
and therefore increase the practicality of the method for real-world applications.
Lastly, we also measure the computational efficiency of all models using FLOPS, inference time (wall clock time), and the number of trainable parameters.
\begin{table}[t]
\centering
\renewcommand{\arraystretch}{1.2} 
\resizebox{\textwidth}{!}{%
\begin{tabular}{@{} l c l 
  S[table-format=2.2] 
  S[table-format=2.2 \pm 2.2, separate-uncertainty=true] 
  S[table-format=2.2 \pm 2.2, separate-uncertainty=true] 
  *{2}{S[table-format=1.2 \pm 2.2, separate-uncertainty=true]} @{}}
\toprule
 &  & \textbf{Model} & {\textbf{FAD ($\downarrow$)}} & {\textbf{DNSMOS ($\uparrow$)}} & {\textbf{SISDRi ($\uparrow$)}} & {\textbf{WER ($\downarrow$)}} & {\textbf{SR-CS ($\uparrow$)}} \\
\midrule
\rowcolor{gray!20}\cellcolor{white} & U & SE-MSB (10)     & 0.35      & 2.72 +- 0.01        & {\textbf{4.95 $\pm$ 0.13}}        & 0.08 +- 0.00          & 0.74 +- 0.00 \\
\rowcolor{gray!20}\cellcolor{white} & U & SE-MSB (50)     & {\textbf{0.33}}      & {\textbf{2.74 $\pm$ 0.01}}        & 4.91 +- 0.13        & 0.09 +- 0.00          & 0.75 +- 0.00 \\ 
\rowcolor{gray!20}\cellcolor{white} \multirow{-3}{*}{Reverb + Noise} & U & BUDDy (100 $\dagger$)  & 0.70      & 2.66 +- 0.01        & 0.73 +- 0.10        & {\underline{\textbf{0.05 $\pm$ 0.00}}}          & {\textbf{0.83 $\pm$ 0.00}} \\
                                    & P & Sepformer  & 0.29    & 2.90 +- 0.01    & 5.00 +- 0.08    & {\underline{0.05 $\pm$ 0.00}}      & 0.83 +- 0.00 \\
                                    & P & SGMSE+    & 0.30    & 2.80 +- 0.01    & 4.49 +- 0.11    & {\underline{0.05 $\pm$ 0.01}}      & 0.84 +- 0.00 \\
                                    & P & UNIVERSE++    & {\underline{0.27}}    & {\underline{3.01 $\pm$ 0.01}}    & {\underline{6.32 $\pm$ 0.09}}    & 0.08 +- 0.00      & {\underline{0.88 $\pm$ 0.00}} \\
\cmidrule{2-8}
                                    &   & Identity     & 1.05   & 1.66 +- 0.03  & 0.00 +- 0.00  & 0.04 +- 0.00    & 0.79 +- 0.00 \\
\midrule
\rowcolor{gray!20}\cellcolor{white} & U & SE-MSB (10)     & {\textbf{0.30}}      & {\textbf{2.76 $\pm$ 0.01}}        & {\textbf{3.47 $\pm$ 0.13}}        & 0.10 +- 0.00          & 0.69 +- 0.01 \\
\rowcolor{gray!20}\cellcolor{white} & U & SE-MSB (50)     & {\textbf{0.30}}      & {\textbf{2.76 $\pm$ 0.01}}        & {\textbf{3.47 $\pm$ 0.13}}        & 0.10 +- 0.00          & 0.69 +- 0.01 \\ 
\rowcolor{gray!20}\cellcolor{white}\multirow{-3}{*}{Reverb + Noise + Clip} & U & BUDDy (100 $\dagger$)  & 0.71      & 2.64 +- 0.01        & 0.79 +- 0.10        & {\underline{\textbf{0.05 $\pm$ 0.00}}}          & {\textbf{0.80 $\pm$ 0.00}} \\
                                    & P & Sepformer  & 0.31    & 2.80 +- 0.01    & 5.16 +- 0.08    & {\underline{0.05 $\pm$ 0.00}}      & 0.79 +- 0.00 \\
                                    & P & SGMSE+    & 0.33    & 2.78 +- 0.01    & 4.64 +- 0.09    & {\underline{0.05 $\pm$ 0.01}}      & 0.80 +- 0.00 \\
                                    & P & UNIVERSE++    & {\underline{0.28}}    & {\underline{3.01 $\pm$ 0.01}}    & {\underline{6.34 $\pm$ 0.08}}    & 0.09 +- 0.00      & {\underline{0.86 $\pm$ 0.00}} \\
\cmidrule{2-8}
                                    &   & Identity     & 1.06   & 1.63 +- 0.03  & 0.00 +- 0.00  & 0.04 +- 0.00    & 0.76 +- 0.00 \\
\midrule
\rowcolor{gray!20}\cellcolor{white} & U & SE-MSB (10)     & 0.48 & 2.25 +- 0.03 & \multicolumn{1}{c}{--} & \multicolumn{1}{c}{--} & \multicolumn{1}{c}{--} \\
\rowcolor{gray!20}\cellcolor{white} & U & SE-MSB (50)     & 0.46 & {\underline{\textbf{2.27 $\pm$ 0.03}}} & \multicolumn{1}{c}{--} & \multicolumn{1}{c}{--} & \multicolumn{1}{c}{--} \\ 
\rowcolor{gray!20}\cellcolor{white}\multirow{-3}{*}{CHiME-6} & U & BUDDy (100 $\dagger$)  & {\underline{\textbf{0.45}}} & 2.16 +- 0.04 & \multicolumn{1}{c}{--} & \multicolumn{1}{c}{--} & \multicolumn{1}{c}{--} \\
                                    & P & Sepformer & 0.90 & 1.19 +- 0.02 & \multicolumn{1}{c}{--} & \multicolumn{1}{c}{--} & \multicolumn{1}{c}{--} \\
                                    & P & SGMSE+  & 0.64 & 1.76 +- 0.04 & \multicolumn{1}{c}{--} & \multicolumn{1}{c}{--} & \multicolumn{1}{c}{--} \\
                                    & P & UNIVERSE++  & 0.66 & 1.87 +- 0.03 & \multicolumn{1}{c}{--} & \multicolumn{1}{c}{--} & \multicolumn{1}{c}{--} \\
\cmidrule{2-8}
                                    &   & Identity     &  1.09 & 1.32 +- 0.02 &  \multicolumn{1}{c}{--} &  \multicolumn{1}{c}{--} &  \multicolumn{1}{c}{--} \\
\bottomrule
\end{tabular}%
}
\caption{
    Evaluation metrics for the mixed speech enhancement tasks, namely WHAM! noise with reverb, and WHAM! noise with reverb and amplitude clipping, with 95\% confidence intervals. 
    The Identity baseline corresponds to unprocessed audio.
    The number of diffusion steps is denoted in parentheses when relevant, except for $\dagger$, which denotes optimization steps.
    U = unpaired, P = paired. Best unpaired model is \textbf{bold}, and best model across paired and unpaired models is \underline{underlined}.
}
\label{tab:degradations}
\end{table}
The generated samples are evaluated using WER (word error rate), 
pMOS/DNSMOS (predicted mean opinion score using the DNSMOS model) \cite{dnsmos}, 
SR-CS (speaker recognition cosine similarity), scale-invariant signal-to-distortion ratio (SISDRi)~\cite{roux2018sdrhalfbakeddone},
and FAD (Fréchet Audio Distance)~\cite{fad}. 
WER, SR-CS, and SISDRi can be interpreted as content fidelity measures capturing how well content is preserved, 
while pMOS and FAD can be interpreted as perceptual quality metrics. 
WER is calculated using the small Whisper model~\cite{whisper} to obtain ground-truth transcriptions of the reference (clean) speech. 
The transcription model is robust towards distorted speech and is therefore not necessarily a good measure of how well the audio is reconstructed. 
Nevertheless, WER serves as a useful indicator of whether the semantic content of the speech is preserved. 
SR-CS is calculated using a speaker embedding model~\cite{titanet} to obtain speaker embeddings of the ground truth signal and the generated signal. 
The metric is then calculated as the cosine similarity between these embeddings. 
The samples used in the FAD metric are encodings produced using a CLAP model~\cite{clap_model}. 
\section{Results}
Results of the dereverberation task are shown in Table \ref{tab:reverberation},
results of the mixed degradation task are shown in Table \ref{tab:degradations}, 
and the results of the few-step sampling experiment are shown in Figure \ref{fig:diffusion_steps}. Computational efficiency is shown in Table \ref{tab:compute}.
The dereverberation and computational efficiency results show that while SE-MSB is moderately outperformed by BUDDy on most metrics, 
SE-MSB is around two orders of magnitude more efficient than BUDDy in terms of FLOPS. This large computation difference is made possible thanks to the linear time sequence scaling of the Mamba diffusion model.
Additionally, we see that while all unpaired methods are outperformed by the paired methods (SGMSE and Paired SE-MSB) on most metrics, the difference is negligible for some metrics, showing that SE-MSB can achieve performance comparable to paired methods.

The results of the mixed degradation task show that SE-MSB outperforms BUDDy by a large margin when the 
degradation is no longer only a convolution with a room impulse response, but also includes additive noise and clipping.
SE-MMSB achieves comparable performance to the supervised Sepformer model on the mixed degradation task.
Finally, the few-step sampling results show that the performance of SE-MSB
is almost unaffected by the number of sampling steps,
while the performance of GFB degrades significantly when using fewer sampling steps.
Interestingly, WER and, to a lesser extent, SISDRi are negatively affected 
when using more sampling SE-MSB, possibly hinting at a perception-distortion trade-off~\cite{perception_distortion}. Because SE-MSB maintains high performance at lower step counts, it further reduces the necessary computational budget required for effective speech enhancement. 
\section{Conclusion}
We proposed an end-to-end unpaired speech enhancement model called SE-MSB, which is a novel combination of Diffusion Schrödinger Bridges and an efficient Mamba Diffusion model architecture operating in the raw waveform domain. 
We demonstrated the utility of the SE-MSB approach in the audio domain, and in particular speech enhancement, for mapping the distribution of degraded speech to the distribution of clean speech from raw waveforms. SE-MSB performs on par with or outperforms other state-of-the-art methods for unpaired speech enhancement, while being orders of magnitude faster during inference.

We also showed that while SE-MSB is moderately outperformed by paired speech enhancement methods, 
the performance gap is, for some metrics, negligible, meaning that SE-MSB is a 
compelling method for adapting speech enhancement models to in-the-wild data without the need for paired examples,
thereby improving the overall performance of speech enhancement models in unknown acoustic environments.

This shows renewed promise for efficient and robust speech enhancement in domains where paired data are unavailable or prohibitively expensive to obtain, such as Lombard speech, style transfer , and the reconstruction of historical recordings.

\newpage
\bibliographystyle{IEEEbib}
\bibliography{refs}

\newpage
\appendix
\section{Training}
\label{appendix:training}
We use the AdamW optimizer \cite{adamw} with a learning rate of $10^{-4}$, $\beta_1=0.9$, $\beta_2=0.999$, and a weight decay of $0.0$.
We use a constant learning rate. 
The models are trained for 250k iterations during pre-training and 100k iterations during fine-tuning.
During fine-tuning, we simulate the forward and backward processes using 20 diffusion steps.
Data is resampled to 16 kHz and chopped/extended to exactly 1.00 seconds during training, 
and exactly 5.12 seconds during testing. All audio is normalized to -20 dBFS.
After normalization, each audio sample is multiplied by 20 to ensure gradients of a reasonable scale during training, 
and we use an L1 loss between the model prediction and the target output.
Gradients are clipped to a maximum norm of 1.0, and we use mixed precision training (BF16). 
Models are compiled using PyTorch's \texttt{torch.compile} for faster training.
We use a batch size of 8, meaning that the effective batch size is 16 since we train the forward and backward processes simultaneously.
We use EMA parameters with a decay of 0.9999 and a warmup period of 89900 steps.
Each SE-MSB model is trained on a single NVIDIA GeForce RTX 4090 GPU with 24GB of VRAM for approximately 1 day and 9 hours.

\subsection{Dataset descriptions}
\label{appendix:datasets}
\subsubsection*{VCTK}
We use the VCTK dataset \cite{vctk} for clean speech samples during training and evaluation.
The dataset consists of 109 English speakers. 
Each speaker utters approximately 400 sentences for a total of approximately 44 hours of speech data.
The audio is recorded at 48 kHz but resampled to 16 kHz during training and evaluation.
We choose to use VCTK specifically due to its prominence in the unpaired speech enhancement literature,
therefore enabling a fairer comparison to other methods. 
The dataset is available online at \href{https://huggingface.co/datasets/badayvedat/VCTK}{huggingface.co/datasets/badayvedat/VCTK}

\subsubsection*{Room Impulse Responses (RIRs)}
For the RIRs, we use a number of different datasets available online \cite{mitreverb, butreverb, openairreverb, rwcpreverb, c4dmreverb}.
We use the same RIR datasets as the GFB model for a fairer comparison. 
In total, the RIR dataset consists of approximately 50 minutes of audio data.
The collection of all RIR datasets is available online at \href{https://huggingface.co/datasets/andnymand/RIR-datasets}{huggingface.co/datasets/andnymand/RIR-datasets}.

\subsubsection*{WSJ0 Hipster Ambient Mixtures (WHAM!)}
For the noise samples, we use the WHAM! dataset \cite{wichern2019whamextendingspeechseparation}.
The WHAM! dataset consists of approximately 78 hours of noise samples recorded in various real-world environments.
The audio is sampled at 48 kHz but resampled to 16 kHz during training and evaluation.
The dataset is available online at \href{https://huggingface.co/datasets/philgzl/wham}{huggingface.co/datasets/philgzl/wham}.

\subsubsection*{Libri Speech}
\label{appendix:libri}
We also evaluate SE-MSB and baselines for the dereverberation task on the LibriSpeech ASR corpus, a large-scale corpus of read English speech~\cite{libri}.
Here, we use only the \texttt{test-clean} split for evaluation with approximately 5.4 hours of clean speech from 40 different speakers. 
The dataset is available online at \href{https://huggingface.co/datasets/openslr/librispeech_asr}{huggingface.co/datasets/openslr/librispeech\_asr}.

\section{Model architecture}
\subsubsection*{Mamba Diffusion Model}
\label{appendix:model_description}
For model architecture, we use a custom Mamba Diffusion Model built on top of the Mamba2 architecture\footnote{https://github.com/state-spaces/mamba} \cite{mamba2}.
The architecture is visualized in figure \ref{fig:model_architecture}.
The model takes as input the raw waveform $x_t$, the current timestep $t\in [0,1]$, and a binary conditioning variable $c$ indicating whether the model is training the forward or backward process.
The waveform $x_t$ is processed by a 1D convolutional layer with a kernel size of 256, a stride of 16, and a padding of 120.
The timestep $t$ is processed by a timestep embedder similar to the one used in \cite{dhariwal2021diffusionmodelsbeatgans}.
The conditioning variable $c$ is encoded using a learned embedding, one for each direction, and added to the timestep embedding.
The "flip" block refers to a flip of the time dimension, and the "concat" block refers to a concatenation along the channel dimension.
Each Mamba block uses a model dimension of 512 and a state space dimension of 128. 
The remaining hyperparameters are the same as the default Mamba2 hyperparameters in the \texttt{mamba-ssm} library.
Our model uses 10 Mamba Diffusion Blocks, and the total number of parameters is approximately 45 million.
The inputs has the following shapes: $x_t\in \mathbb{R}^{B \times C \times T}$, $t \in \mathbb{R}^{B}$, and $c \in \{0,1\}^{B}$,
where $B$ is the batch size, $C$ is the number of channels (1 in all our cases), and $T$ is the sequence length.
The output of the model has the same shape as the input waveform $x_t$.
The input to the Mamba Diffusion Block, $x$ and $c$, has shape $x \in \mathbb{R}^{B \times C' \times T'}$ and $c \in \mathbb{R}^{B}$,
where $C'$ and $T'$ are the number of channels and the sequence length after the initial convolutional layer, respectively.
For our specific convolutional layer, $C'=512$ and $T' = T/16$.
Here, $c$ is the conditioning variable consisting of the embeddings of the current timesteps $t$ and the indicator variable $s$,
and $x$ is the output of the initial convolutional layer or the output of the previous Mamba Diffusion Block.

\section{Baselines}
\label{appendix:baselines}
\subsubsection*{Gaussian Flow Bridge (GFB)}
We use the GFB implementation provided by the original authors \cite{moliner2024gaussianflowbridgesaudio}.
Their code is available at \href{https://github.com/microsoft/GFB-audio-control}{github.com/microsoft/GFB-audio-control}.
A Gaussian Flow Bridge is an unsupervised generative model that learns to map two or more distributions to a shared latent space,
namely a Gaussian distribution. 
Then, during inference, the model can map samples to and from the latent space, 
therefore enabling transformations between the distributions that the model was trained on.
We use the pre-trained model checkpoint for reverberant speech enhancement with chunk size $N_c=128$ available on the GitHub repository. 
For more info on the chunk size, please refer to their paper and codebase.
The GFB model is also trained on clean speech from the VCTK dataset with the same reverberation datasets as our model, 
also at 16 kHz. 
The GFB model uses approximately 44 million parameters, which is comparable to our model's 45 million parameters.
Importantly, the GFB model is conditioned on so-called reverberant descriptors, namely reverberation time ($T_{60}$) and
clarity ($C_{50}$). These two descriptors are explicitly provided to the model as a conditioning variable.
Our model is not conditioned on any explicit descriptors or any other auxiliary information about the degradation.

\subsubsection*{Blind Unsupervised Dereverberation with Diffusion Models (BUDDy)}
We use the BUDDy implementation provided by the original authors \cite{buddy}.
Their code is available at \href{https://github.com/sp-uhh/buddy}{github.com/sp-uhh/buddy}.
BUDDy is an unsupervised dereverberation method that uses a pre-trained diffusion model as a prior for the clean speech distribution.
It works by jointly estimating the acoustic parameters of the reverberation and the clean speech signal. 
We use the pre-trained model checkpoint available \href{https://drive.google.com/uc?id=1j-BHDqiKxbEfpDUpc1GHZXeoHIA1nIal}{online}.
The pre-trained diffusion model uses the NCSN++M model architecture \cite{storm} with approximately 27.8 million parameters. 
The model is also trained on clean speech from the VCTK dataset at 16 kHz. 
BUDDy is not explicitly trained for dereverberation, but rather the model assumes that the input is clean speech
convolved with the acoustic characteristics of the room.

\subsubsection*{Weighted Prediction Error (WPE)}
We use the NARA-WPE implementation provided by \cite{Drude2018NaraWPE}.
The WPE algorithm models late reverberation as a delayed linear autoregressive process of previously observed signals in the STFT domain. The main assumption is that each current audio frame can be split in two parts, one part for desired early speech and one part for undesired late reverberation.
WPE estimates the late reverberant tail by applying a linear prediction filter to past frames. It is a maximum likelihood method that requires no prior knowledge of the room acoustics.
We use the following hyperparameters for the WPE algorithm:
\texttt{taps=20, delay=3, iterations=5, stft\_size=512, stft\_shift=128}, and \texttt{statistics\_mode="full"}.

\subsection*{Score-based Generative Models for Speech Enhancement (SGMSE)}
We use the SGMSE implementation provided by the original authors \cite{sgmse}. 
Their code is available at \href{https://github.com/sp-uhh/sgmse}{github.com/sp-uhh/sgmse}.
The model is trained on clean speech samples from the WSJ0 dataset \cite{WSJ0}, 
and each clean speech sample is convolved with a simulated room impulse response.
SGMSE is trained for paired dereverberation.
We use their pre-trained model checkpoint available \href{https://drive.google.com/uc?id=1eiOy0VjHh9V9ZUFTxu1Pq2w19izl9ejD}{online}.
The model uses the NCSN++ model architecture \cite{song2021scorebasedgenerativemodelingstochastic} with 
approximately 65 million parameters.

\subsubsection*{Paired SE-MSB / Audio-to-Audio Schrödinger Bridge (A$^2$ASB)}
For the A$^2$ASB baseline, in the paper called Paired SE-MSB, we adapt a different implementation compared to the original authors \cite{kong2025a2sbaudiotoaudioschrodingerbridges}.
The original A$^2$ASB implementation is trained for bandwidth extension and inpainting on 44 kHz audio using an adapted frequency domain representation.
Instead, we train our own A$^2$ASB model for dereverberation using the same architecture,
data, and parameters as the unpaired SE-MSB model, see section \ref{appendix:training}.
Theoretically, the only difference between SE-MSB and A$^2$ASB is that the latter is trained on paired examples, in addition to the network architecture and data representation.
Generally, A$^2$ASB can be seen as the paired version of the SE-MSB model.
Formally, A$^2$ASB only trains the backward process, and intermediate samples are drawn from the distribution:
\[
\Xt{k}\sim \mathcal{N}(\mu_t, \Sigma_t)
\]
Where 
\[
\mu_t = \frac{\bar{\sigma}_t^2 + \X_0 \sigma_t^2\X_1}{\bar{\sigma}_t^2 + \sigma_t^2},\hspace{5mm} \Sigma_t = \frac{\bar{\sigma}_t^2\sigma_t^2\mathbf{I}}{\bar{\sigma}_t^2 + \sigma_t^2}
\]
and
\[
\sigma_t^2 = \int_0^t\beta_{\tau} d\tau, \hspace{5mm} \bar{\sigma}_t^2 = \int_t^1\beta_{\tau} d\tau
\]
Assuming that $\beta_t = \beta$ is constant, we can simplify the above to:
\[
\mu_t = (1-t)\X_0 + t\X_1,\hspace{5mm} \Sigma_t = \beta t(1-t)\mathbf{I}
\]
This formulation is mathematically equivalent to the DSB Matching intermediate distribution in \ref{eq:intermediate}.
Similarly, A$^2$ASB uses the following posterior distribution for the backward process:
\[
p(\Xt{t-\Delta t} | \X_0, \X_t) = \mathcal{N}\left( \frac{(\Delta \sigma_t^2)\X_0 + \sigma_t^2 \X_t}{\Delta \sigma_t^2 + \sigma_t^2}, \frac{(\Delta \sigma_t^2)\sigma_t^2}{\Delta \sigma_t^2 + \sigma_t^2}\mathbf{I} \right)
\]
Where $\Delta \sigma_t^2 = \sigma_t^2 - \sigma_{t - \Delta t}^2$.
This is mathematically equivalent to the posterior distribution in \ref{eq:p_backward} assuming a constant $\beta_t$.
 
\subsubsection*{Sepformer (Separation Transformer)}
We use a SepFormer model implemented with speechbrain available at \href{https://huggingface.co/speechbrain/sepformer-wham16k-enhancement}{huggingface.co}. The SepFormer is a transformer based neural network for speech seperation, that learns short and long-term dependencies through a learned encoded representation and iterative masking procedure done by two Multi-Head Attention blocks called an IntraTransformer and InterTransformer respectively. The model is trained on the \textit{WHAM!} dataset, which we also evaluate on for the mixed speech enhancment tasks. The SepFormer is trained for denoising, and not mixed tasks.

\section{Computational efficiency}
\label{appendix:efficiency}
In addition to TFLOPS, we also report inference time in milliseconds for all dereverberation models and the number of trainable parameters in Figure \ref{tab:compute}.
TFLOPS vs Inference time is based on 100 forward passes on samples of 2.048 seconds at 16 kHz using a batch size of 1 and inference mode. 
Inference time is measured on a consumer-grade GPU and CPU, namely an AMD Ryzen 7 7700 8-Core Processor CPU with 16GB of RAM, and an NVIDIA RTX 4070 Super with 12GB of VRAM. 
To show that the selective scan of the Mamba2 block has a flop count of $FLOP_{scan} = 2 B T  N(H\ Q + 3 D)$, we refer to chapter 6 of the original Mamba2 paper \cite{mamba2}. They introduce the notation $BMM(B,M,N,K)$ to define a batched matrix multiplication with a total computation cost of $O(BMNK)$ FLOPs per batch element per head. The $O()$ notation here signifies a multiply and an add function, so we use a factor $2$. According to the paper the computational cost arises from a kernel matrix computation with cost $BMM(T/Q,Q,Q,N) = 2 \cdot B\cdot T \cdot Q \cdot N \cdot H$, an input state multiplication with a cost of $BMM(T/Q,Q,P,N) = 2\cdot B \cdot T \cdot H \cdot P \cdot N$. There are then three low rank block processes, one right factor with a cost of $BMM(T/Q,N,P,Q) = 2 \cdot B \cdot T \cdot H \cdot P \cdot N$, a left factor with a cost of $BMM(T/Q,Q,P,N)$ with the same cost as the right factor, and a central factor with a cost that is noted negligible by the authors. Summing, we get a total cost of
$$
FLOP_{scan} = 2\cdot B \cdot T \cdot N\left(Q  \cdot H + 3\cdot H \cdot P \right) = 2\cdot B \cdot T \cdot N\left(Q  \cdot H + 3D\right),
$$
where it has been used that $D = H\cdot P$ is the total inner dimension, resulting from a head dimension $P$ times number of heads $H$.

\subsubsection*{WPE Flops}
For WPE, the STFT-transformed input signal is split into $F$ frequency bins over $T$ time frames. The signal is split into $D_{c}$ microphone channels and $K_{t}$ filter taps. The algorithm requires calculating a covariance matrix $R_f$ by an outer product of two  $D_cK_t \times 1$ complex vectors for each frequency bin and time frame, requiring $8 T F (D_c K_t)^2$ FLOPs. Next, a cross-correlation vector $P_f$ is calculated from a complex $D_cK_t \times 1$ vector, which is multiplied by a complex $D_c \times 1$ vector, requiring $8 T F D_c^2K_t$ FLOPs. These are used to solve a linear system $G_f = R_f^{-1}P_f$, requiring $F  \left( \frac{8}{3}(D_cK_t)^3 + 8(D_cK_t)^2 D_c \right)$ FLOPs, and finally this system is applied to the reverberation tail for an additional $8  F T  D_c^2K_t$ FLOPs.


\end{document}